\documentclass[fleqn,twoside]{article}
\usepackage{espcrc2}
\usepackage{epsfig}

\newcommand{\de}{\Delta E}
\newcommand{\mbc}{M_{\rm bc}}
\newcommand{\mds}{M(D_s)}
\newcommand{\bdsk}{{\bar B^0}\to D_s^+K^-}
\newcommand{\bdspi}{B^0\to D_s^+\pi^-}
\newcommand{\bdppi}{{\bar B^0}\to D^+\pi^-}
\newcommand{\bdkstar}{B^-\to D^0 K^{*-}}
\newcommand{\kpi}{K^-\pi^+}
\newcommand{\kpipin}{\kpi\pi^0}
\newcommand{\kpipipi}{\kpi\pi^-\pi^+}
\newcommand{\bdnpin}{\bar{B}^0\to D^0\pi^0}

\title{Open charm $B$ decays}

\author{P. Krokovny\address[BINP]{Budker Institute of Nuclear 
    Physics, Novosibirsk, Russia}, for the Belle Collaboration}
\begin{document}

\begin{abstract}
New results on $B$ decays to charm-related modes from the Belle
and CLEO experiments are presented.
\vspace{1pc}
\end{abstract}

\maketitle

\section{Introduction}
Studies of charm-related $B$ decays provide information about the
dynamics of $B$ meson decays. Results of these fields can be used
to check predictions of theoretical models. Recently new results appeared 
from the $B$ factories at KEK and SLAC and from CLEO. 
This report covers the recent results on this subject from
Belle~\cite{NIM} and CLEO~\cite{CLEO}.

The CLEO results are based on a 9.15~fb$^{-1}$ data sample
collected at the center-of-mass (CM) energy of the $\Upsilon(4S)$ 
resonance, while the Belle results are obtained using various data 
samples from 29.2~fb$^{-1}$ to 78.7~fb$^{-1}$.
Both group identify $B$ candidates by two kinematic variables:
the energy difference, \mbox{$\de=(\sum_iE_i)-E_b$}, and the
beam constrained mass, $\mbc=\sqrt{E_b^2-(\sum_i\vec{p}_i)^2}$, where
$E_b=\sqrt{s}/2$ is the beam energy and $\vec{p}_i$ and $E_i$ are the 
momenta and energies of the decay products of the $B$ meson in the 
CM frame. 

\section{$B\to D\pi$ Isospin Analysis (CLEO Collaboration)}
Precise measurements of the $B\to D\pi$ branching fractions can
be used to extract of the strong phase difference $\delta_I$
between the $I=1/2$ and $I=3/2$ isospin amplitudes in the $D\pi$ system.
Observation of the color-suppressed $\bdnpin$
decay~\cite{color_belle,color_cleo} completed the measurements of the
$D\pi$ final states. Recently CLEO collaboration improved the accuracy
in the $B^-\to D^0\pi^-$ and $\bdppi$ branching 
fractions~\cite{dpi_cleo}, these results are presented here.

\begin{figure}
  \includegraphics[width=0.45\textwidth] {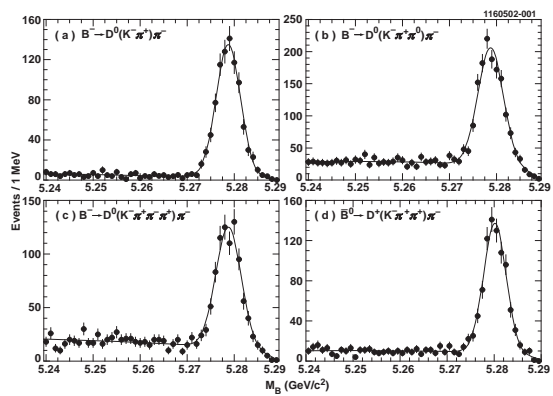}
\vspace*{-5mm}
  \caption{The $\mbc$ distributions for the $B\to D\pi$ candidates.}
  \label{bdpi_mbc}
\end{figure}

$D^0$ mesons are reconstructed using three decay channels 
$D^0\to\kpi$, $\kpipipi$ and $\kpipin$.
Charged $D$ mesons are reconstructed via the mode $K^-\pi^+\pi^+$.
In each case, $D$ candidates are required to have an invariant mass 
within $3\sigma$ of the nominal $D$ mass.
The fitted $\mbc$ distributions for each of the $D$ decay modes are
presented in Fig.~\ref{bdpi_mbc}. 
The following branching fractions have been obtained:
${\cal B}(B^-\to D^0\pi^-)=(4.97\pm 0.12\pm 0.29\pm 0.22)
\times 10^{-3}$ and
${\cal B}(\bar{B}^0\to D^+\pi^-)=(2.68\pm 0.12\pm 0.24\pm 0.12)
\times 10^{-3}$.
Here the first error is statistical, the second is systematic and the
third one is a separate systematic error due to the experimental
uncertainty of the production fractions of charged and neutral $B$
mesons from $\Upsilon(4S)$ decays. Using the measurements of 
${\cal B}(\bar{B}^0\to D^0\pi^0)$~\cite{color_belle,color_cleo},
$\cos \delta_I=0.863^{+0.024+0.036+0.038}_{-0.023-0.035-0.030}$
has been obtained. The difference of $\delta_I$ from zero 
is $2.3\sigma$ indicating the presence of final state interactions 
in $B\to D\pi$ decays.

\section{$B\to D_s \pi/K$ (Belle Collaboration)}

\begin{table*}
\caption{The signal yields and branching fractions
for the $\bdsk$ and $\bdspi$ decay channels.}
\footnotesize
\medskip
\label{dsh_res}
  \begin{tabular*}{\textwidth}{l@{\extracolsep{\fill}}ccccc}\hline\hline
 Mode  & $\mds$ - $\de$ yield & $\mds$ yield & $\de$ yield &
${\cal B}$ $(10^{-5})$ &
Significance\\\hline
$\bdsk$ & $16.4^{+4.6}_{-3.9}$ &
        $15.0^{+4.5}_{-3.8}$ & $17.5^{+4.8}_{-4.2}$ & 
        $4.6^{+1.2}_{-1.1}\pm 1.3$ & $6.4\sigma$\\

$\bdspi$ & $10.1^{+4.4}_{-3.7}$ &
        $10.3^{+4.1}_{-3.4}$ & $9.5^{+4.5}_{-3.8}$ & 
        $2.4^{+1.0}_{-0.8}\pm 0.7$ & $3.6\sigma$\\\hline\hline
  \end{tabular*}
\end{table*}

The decay $\bdspi$ is expected to be dominated by a $b\to u$
transition, with no penguin contribution. Therefore, it can
provide a way to determine the CKM matrix element, $|V_{ub}|$~\cite{vub2}.
The decay $\bdsk$ can occur via $W$-exchange or final state
rescattering and cannot be described by a spectator graph.
The measurement of this decay mode can be used to 
estimate $W$-exchange or final state rescattering contributions
in other $B$ decays.
Only upper limits have been reported for these decays
by CLEO~\cite{dspi_cleo}.
Recently, the BaBar collaboration presented evidence
for the $\bdspi$ decay~\cite{dspi_babar}.

\begin{figure}
  \includegraphics[width=0.45\textwidth] {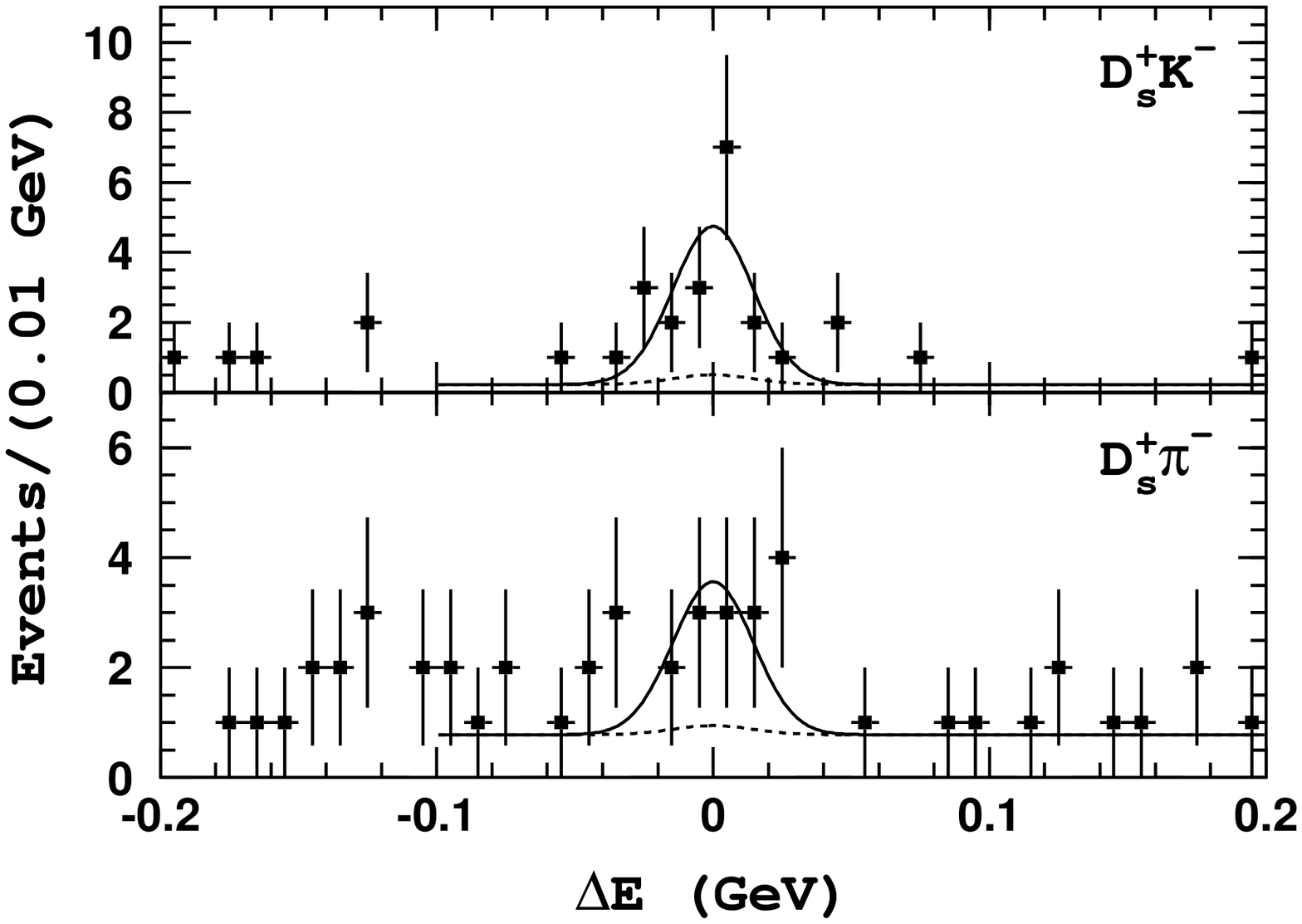}
\vspace*{-1cm}
  \caption{The $\de$ spectra for the $\bdsk$ (top) and $\bdspi$ (bottom)
    candidates. }
  \label{dsh_de}
\end{figure}
\begin{figure}
  \includegraphics[width=0.45\textwidth] {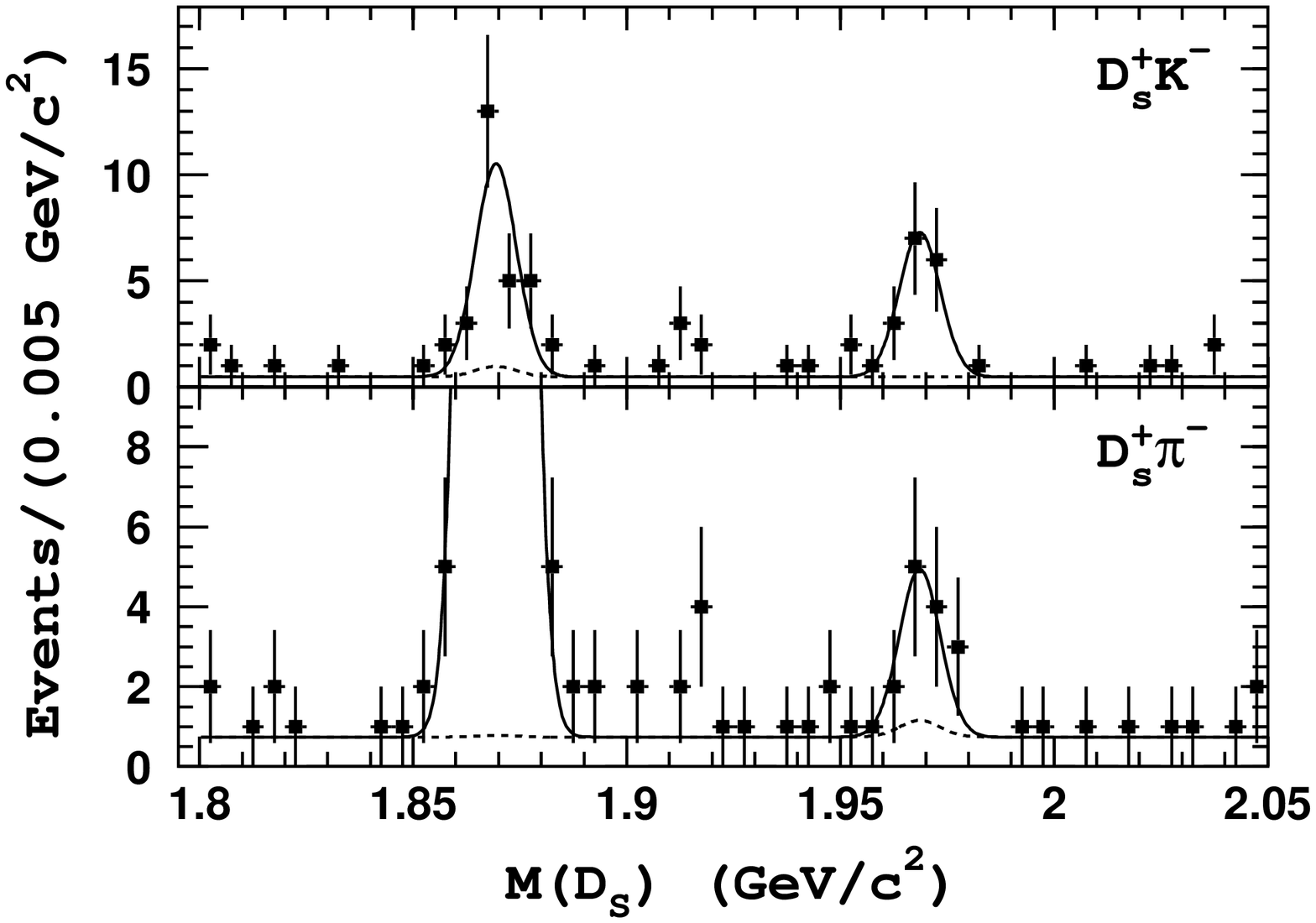}
\vspace*{-1cm}
  \caption{The $\mds$ spectra for the $\bdsk$ (top) and
    $\bdspi$ (bottom) candidates. }
  \label{dsh_mds}
\end{figure}

Here Belle results~\cite{dsh_belle} on a search for these decays are
presented.
The $D_s^+$ candidates are reconstructed in the $D_s^+\to \phi\pi^+$, 
$\bar{K}^{*0}K^+$ and $K_S^0K^+$ decay channels. $D_s^+$ candidates are
combined with a charged kaon or pion to form a $B$ meson.
To extract the signal, a binned maximum likelihood fit to the 
two-dimensional distribution of data in $\mds$ and $\de$ is performed.
The $D_s^+$ signal is described by a two-dimensional Gaussian, while the
background function includes three components: combinatorial (flat in
$\mds$ and $\de$), $q\bar{q}$ events that peak in $\mds$ and are flat in
$\de$, and $B$ decay events that peak in $\de$ and are flat in $\mds$.
The levels of three background components are allowed to vary
independently in the three reconstructed $D_s^+$ modes.

Figures~\ref{dsh_de} and \ref{dsh_mds} show the $\de$ and $\mds$
projections for events from the signal region, the fitted
signal plus background combined shape by solid lines and
background shape including the peaking background by dashed lines.
In addition to the clear signals at the $D_s^+$ mass in
Fig.~\ref{dsh_mds}, also seen are peaks at the $D^+$ mass,
corresponding to the $\bdppi$ and ${\bar B^0}\to D^+ K^-$ decays.
The fit results are given in Table~\ref{dsh_res}.
The results of one-dimensional fits to the $\mds$ and $\de$
distributions are also shown in Table~\ref{dsh_res} for comparison.
A statistically significant signal ($6.4\sigma$) is observed for the
$\bdsk$ decay channel, while only evidence is reported for the
$\bdspi$ decay.

\section{$B^-\to D^{(*)+}\pi^-\pi^-$ (Belle Collaboration)}

\begin{table*}
\caption{The branching fractions and resonance parameters
for the $D^{(*)+}\pi^-\pi^-$ final states.}
\footnotesize
\medskip
\label{d2s_res}
  \begin{tabular*}{\textwidth}{l@{\extracolsep{\fill}}ccc}\hline\hline
 Mode  & ${\cal B}(B^-\to D_X\pi^-){\cal B}(D_X\to D^{(*)+}\pi^-)$ & 
$M(D_X)$, MeV$/c^2$ & $\Gamma(D_X)$, MeV$/c^2$ \\\hline
$B^-\to D_2^{*0}\pi^-\to D^+\pi^-\pi^-$ & 
  $(3.5 \pm 0.3 \pm 0.5)\times 10^{-4}$ & 
  $2460.7\pm 2.1\pm 3.1$ & $46.4\pm 4.4\pm 3.1$\\
$B^-\to D_0^{*0}\pi^-\to D^+\pi^-\pi^-$ & 
  $(5.5 \pm 0.5 \pm 0.8)\times 10^{-4}$ & 
  $2290\pm 22\pm 20$ & $300\pm 30\pm 30$\\
$B^-\to D_v^{*0}\pi^-\to D^+\pi^-\pi^-$ & 
  $(1.4 \pm 0.3 \pm 0.2)\times 10^{-4}$ & \cite{PDG} & \cite{PDG}\\
\hline
$B^-\to D_1\pi^-\to D^{*+}\pi^-\pi^-$ & 
  $(6.2 \pm 0.5 \pm 1.1)\times 10^{-4}$ & 
  $2423.9\pm 1.7\pm 0.2$ & $26.7\pm 3.1\pm 2.2$\\
$B^-\to D_2^{*0}\pi^-\to D^{*+}\pi^-\pi^-$ & 
  $(2.0 \pm 0.3 \pm 0.5)\times 10^{-4}$ & 
  \cite{d2s_note} & \cite{d2s_note}\\
$B^-\to D_1^{*0}\pi^-\to D^{*+}\pi^-\pi^-$ & 
  $(4.1 \pm 0.5 \pm 0.8)\times 10^{-4}$ & 
  $2400\pm 30\pm 20$ & $380\pm 100\pm 100$\\
\hline\hline
  \end{tabular*}
\end{table*}

A study of charmed meson production in $B$ decays provides an
opportunity to test predictions of Heavy Quark Effective Theory (HQET) 
and QCD sum rules. $B$ decays to $D^{(*)}\pi$ final states are the 
dominant hadronic $B$ decay modes and are measured quite well~\cite{PDG}.
The large data sample accumulated in the Belle experiment allows to study  
production of $D$ meson exited states.
$D^{**}$s are P-wave excitations of quark-antiquark  systems that
contain one charmed and one light ($u$,$d$) quark.
The $B\to D^{**}\pi$ decays have been studied by Belle~\cite{d2s_belle}
using the $D^+\pi^-\pi^-$ and the $D^{*+}\pi^-\pi^-$ final states.

\begin{figure}
  \includegraphics[width=0.23\textwidth] {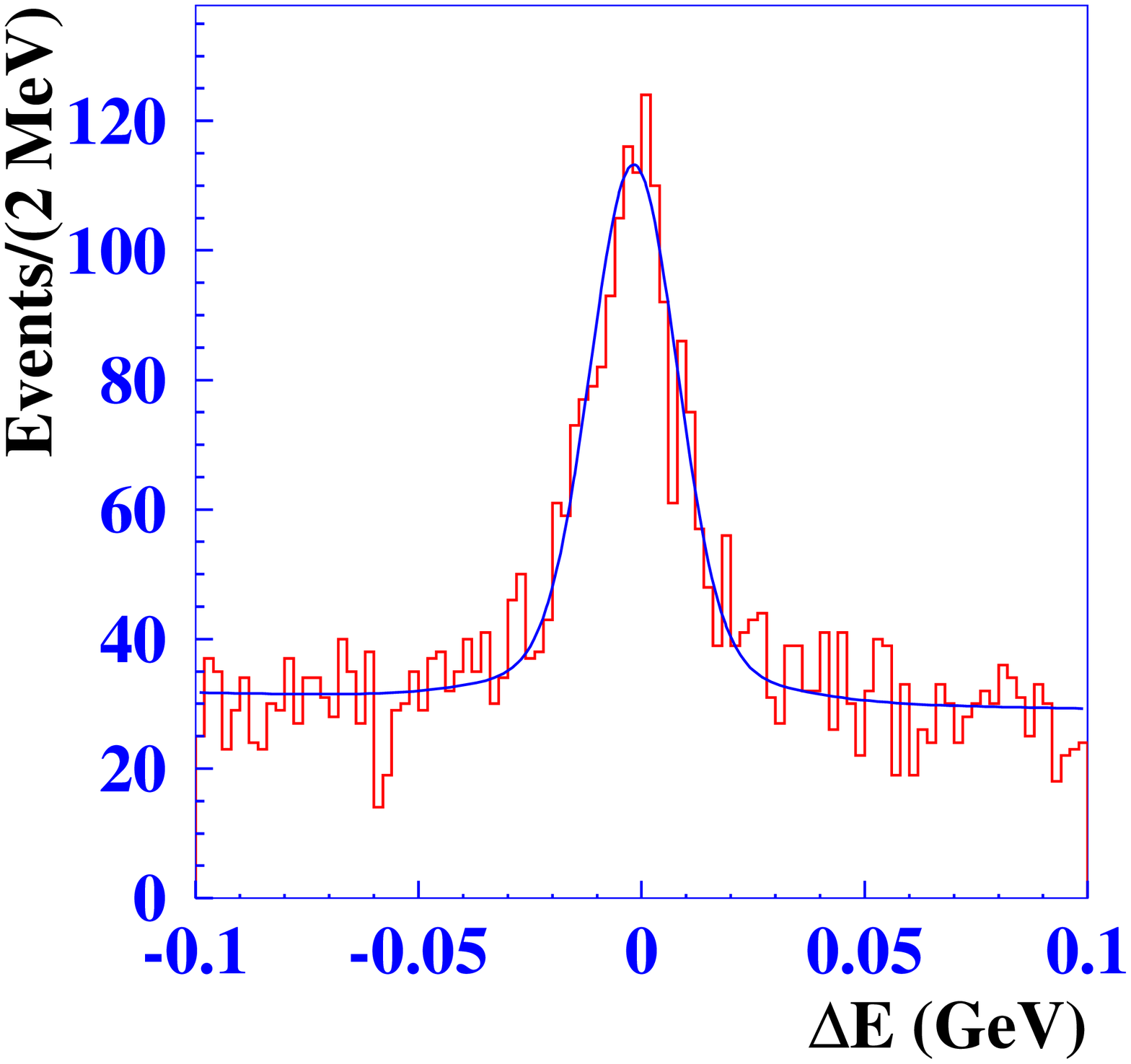}
  \includegraphics[width=0.23\textwidth] {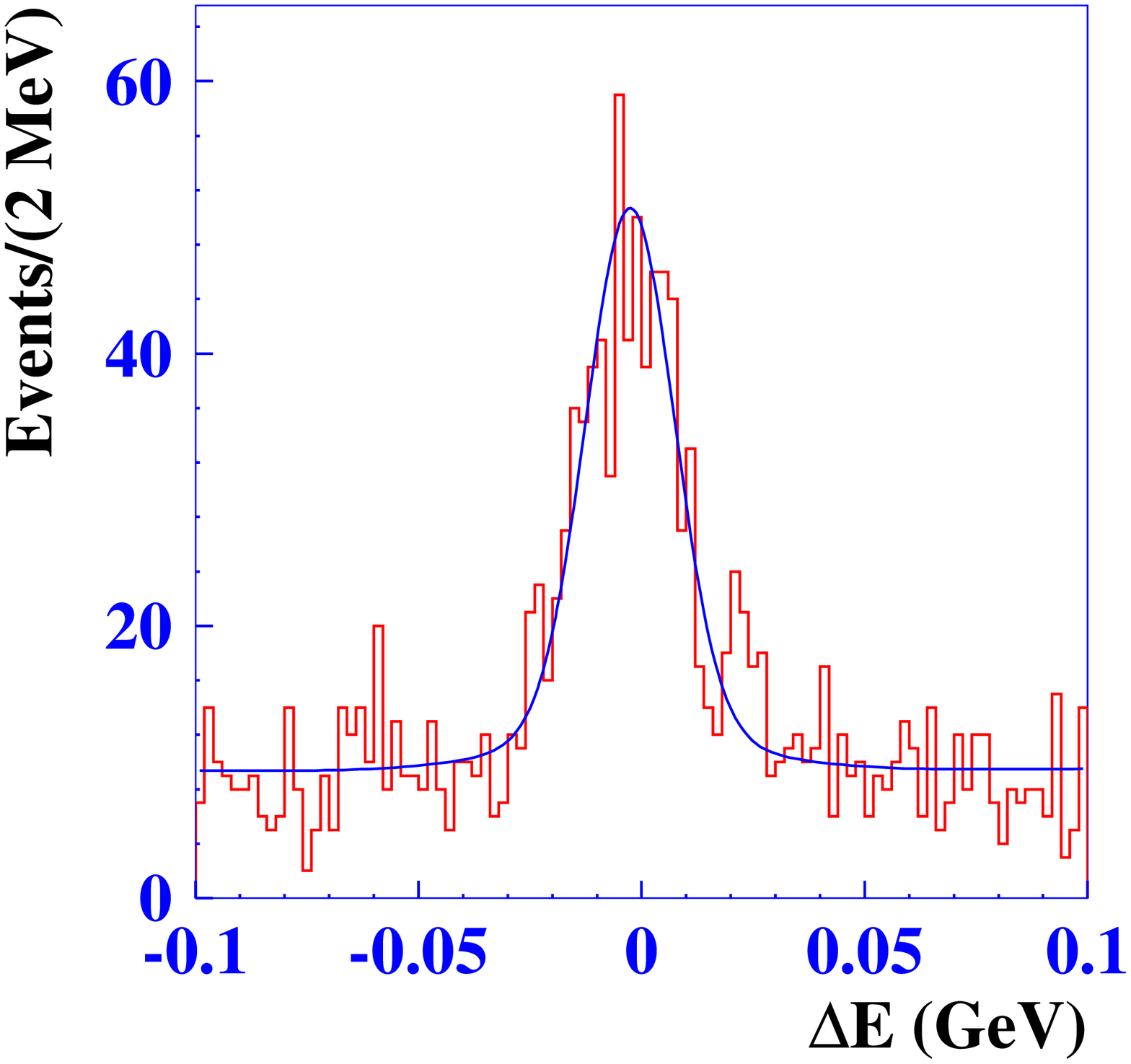}
\vspace*{-1cm}
  \caption{The $\de$ distributions for the $B^-\to D^+\pi^-\pi^-$
  (left) and $B^-\to D^{*+}\pi^-\pi^-$ (right) candidates.} 
  \label{d2s_de}
\end{figure}

Figure~\ref{d2s_de} shows the $\de$ distributions for the 
$B^-\to D^+\pi^-\pi^-$ and $B^-\to D^{*+}\pi^-\pi^-$ candidates.
The following branching fractions are measured:
${\cal B}(B^-\to D^+\pi^-\pi^-)=(1.07\pm0.05\pm0.16)\times10^{-3}$ and
${\cal B}(B^-\to D^{*+}\pi^-\pi^-)=(1.24\pm0.07\pm0.22)\times10^{-3}$,
without any assumption about the intermediate final states.

\begin{figure}
  \includegraphics[width=0.23\textwidth] {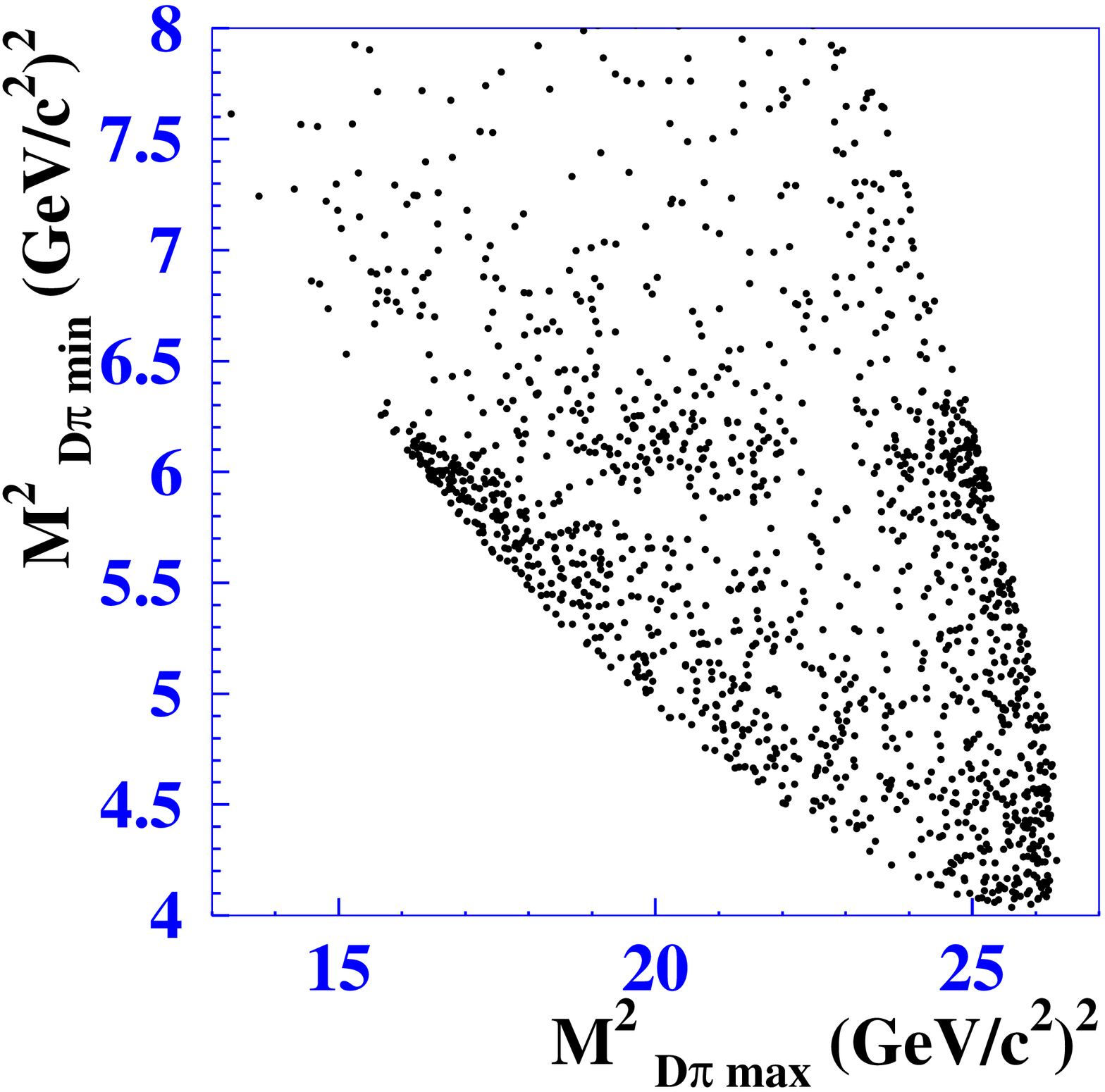}
  \includegraphics[width=0.23\textwidth] {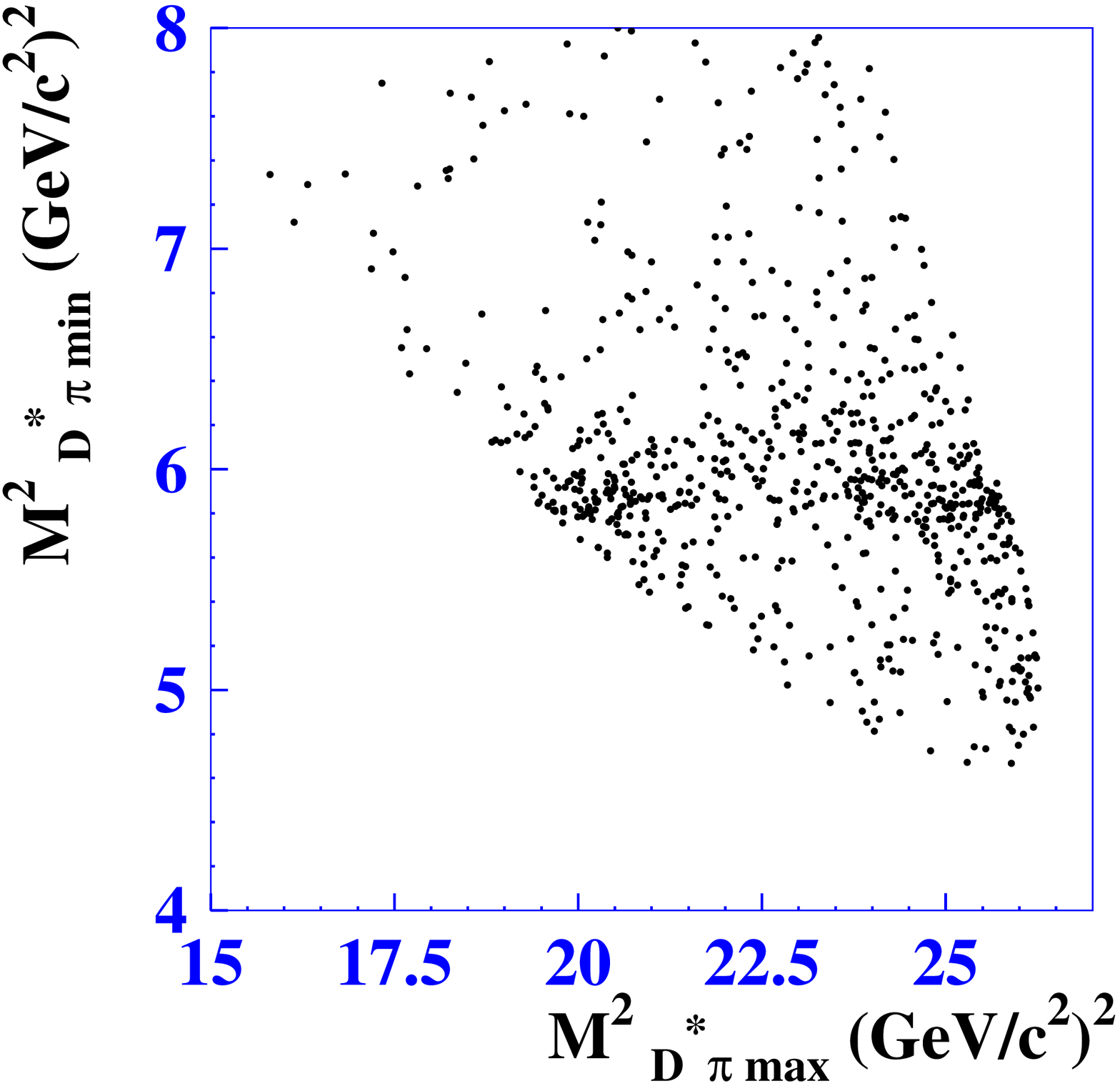}
\vspace*{-1cm}
  \caption{The Dalitz plot distributions for $B^-\to D^+\pi^-\pi^-$
  (left) and $B^-\to D^{*+}\pi^-\pi^-$ (right) candidates.} 
  \label{d2s_dalitz}
\end{figure}

To study the dynamics of $B\to D^{(*)}\pi\pi$ decays, an analysis of 
the Dalitz plots shown in Fig.~\ref{d2s_dalitz} is performed.
The fit to the $D^+\pi^-\pi^-$ Dalitz plot includes three final states:
$D_2^{*0}\pi^-$, $D_0^{*0}\pi^-$ and the contribution of the process with 
virtual $D^{*0}\pi$ production($D_v^{*0}\pi^-$).
The results are presented in Table~\ref{d2s_res}.

Two additional degrees of freedom should be taken into account in the
$D^{*+}\pi^-\pi^-$ final state: the angle ($\alpha$) between the pion 
from the $D^{**}$ decay and the pion from the $D^*$ decay in the $D^*$ 
rest frame and the azimuthal angle ($\gamma$) of this particle
relative to the plane of the $B^-\to D^{*+}\pi^-\pi^-$ decay.
The fit to the $D^{*+}\pi^-\pi^-$ distribution includes three final 
states: $D_1\pi^-$, $D_2^{*0}\pi^-$ and $D_1^{*0}\pi^-$.
The results are presented in Table~\ref{d2s_res}.

Using these measurements the ratio of $D_2^{*0}$ branching
fractions $h={\cal B}(D_2^{*0}\to D^+\pi^-)/
{\cal B}(D_2^{*0}\to D^{*+}\pi^-)=1.77\pm 0.49$, 
consistent with the world average $h=2.3\pm 0.6$~\cite{PDG}, is obtained.
The measured ratio $R={\cal B}(B^-\to D_2^{*0}\pi^-)/
{\cal B}(B^-\to D_1^0\pi^-)=0.89\pm 0.14$ is lower than the CLEO
measurement $1.8\pm 0.8$~\cite{r_cleo} (although the results are
consistent within errors) but is still a factor of 3 larger than the
factorization prediction~\cite{neubert}. 
Belle measurements show that the narrow resonances  
compose  $(33\pm 4)\%$ of the
$D\pi\pi$ decays and $(66\pm 7)\%$ of the $D^*\pi\pi$ decays.
This  result is inconsistent with the QCD sum rule prediction and may
indicate a large contribution from a color suppressed amplitude.

\section{$\bar{B}^0\to D^{(*)0}\pi^+\pi^-$ (Belle Collaboration)}

\begin{figure}
  \includegraphics[width=0.23\textwidth] {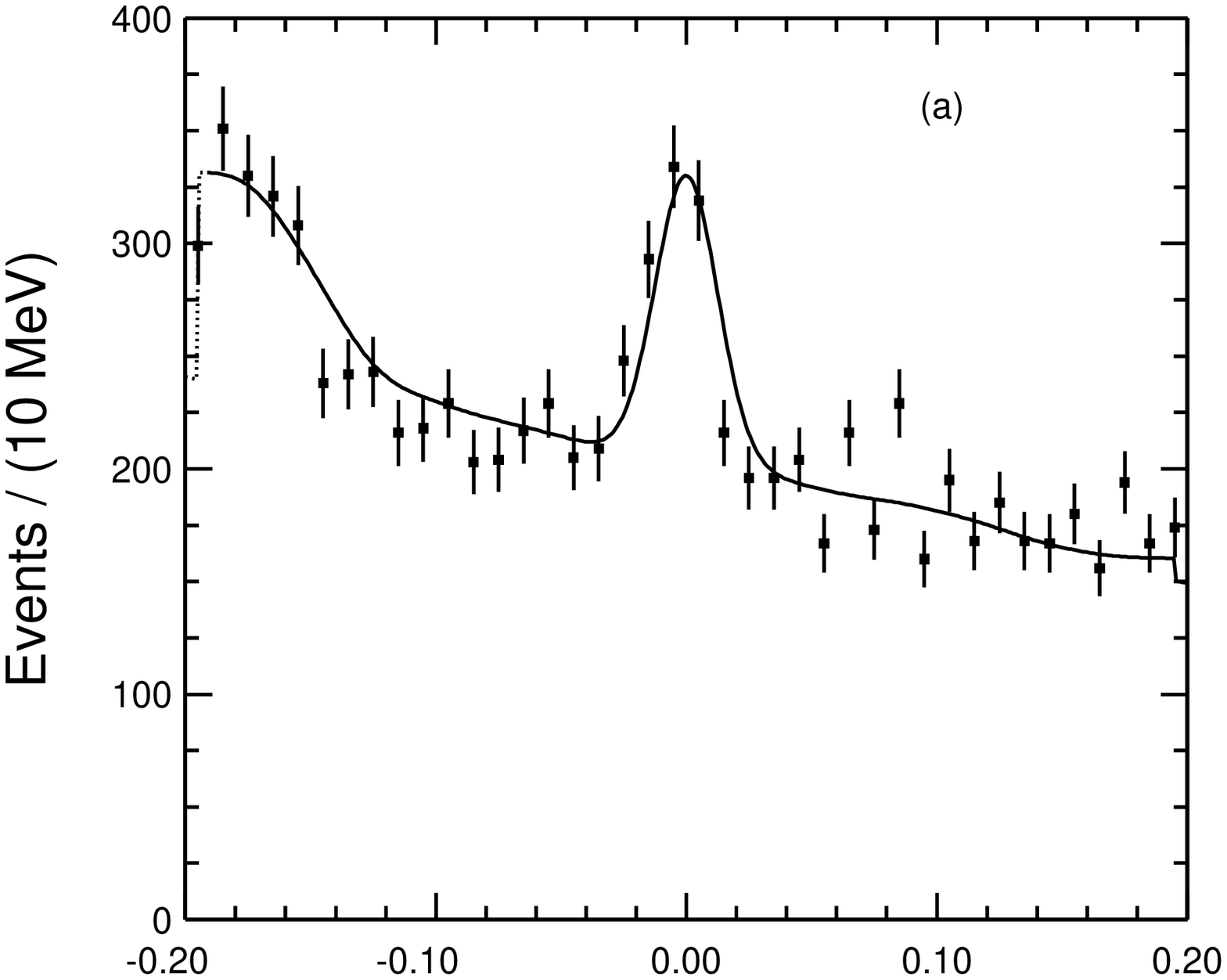}
  \includegraphics[width=0.23\textwidth] {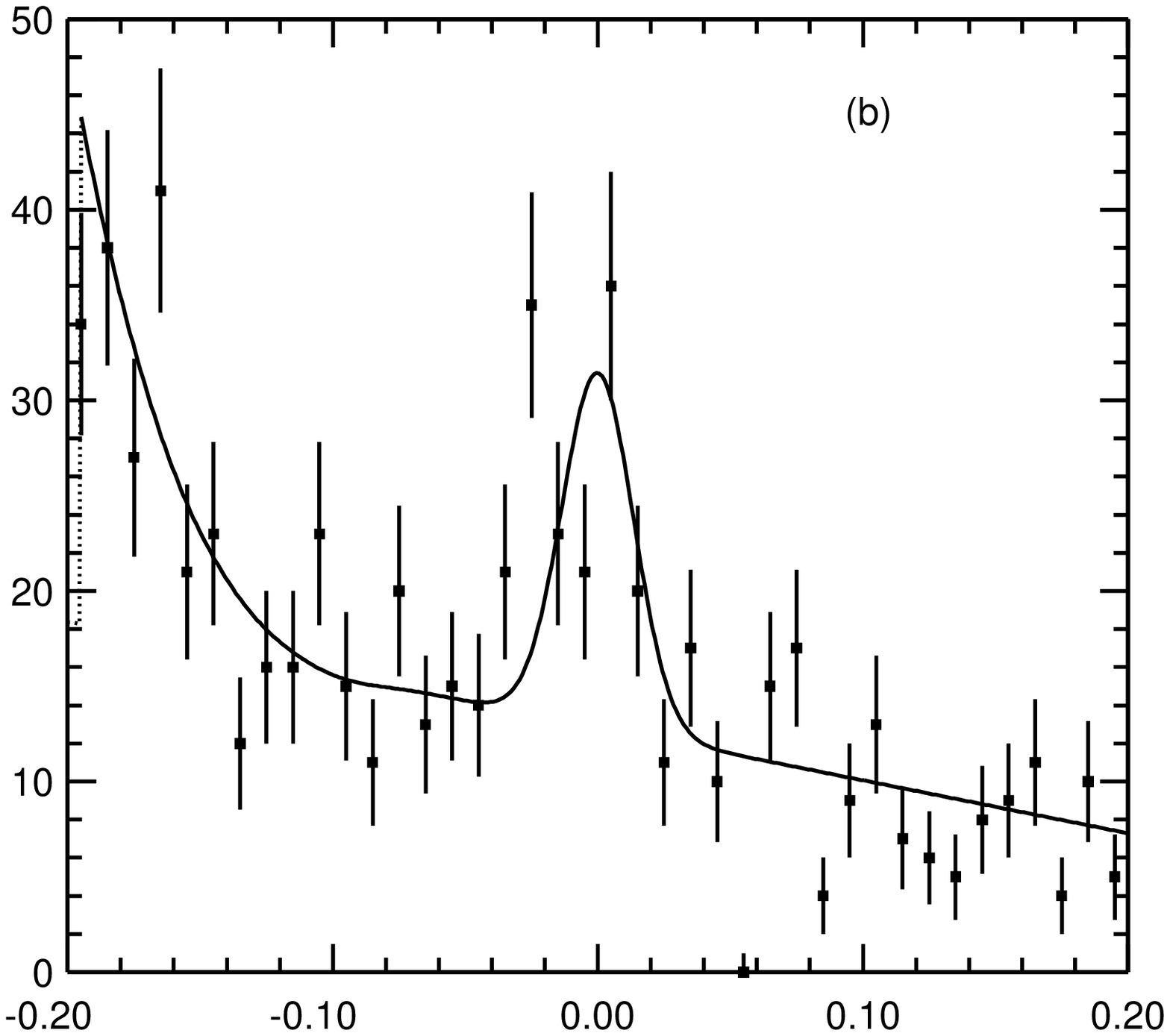}
\vspace*{-1cm}
  \caption{The $\de$ distributions for  (a) 
    $\bar{B}^0\to D^0\pi^+\pi^-$
   and (b) $\bar{B}^{*0}\to D^0\pi^+\pi^-$  candidates.} 
  \label{dpipi_de}
\end{figure}

Exclusive hadronic decay rates provide important tests of models for
$B$ meson decays. The $\bar{B}^0\to D^{(*)0}\pi^+\pi^-$ decays
provide a precision testing ground for factorization~\cite{reader} and 
the possibility to search for resonant substructure in the final state.
At present, only an upper limit ${\cal B}(\bar{B}^0\to
D^0\pi^+\pi^-)<1.6\times 10^{-3}$~\cite{d0pipi_cleo} exists.

\begin{figure}
  \includegraphics[width=0.23\textwidth] {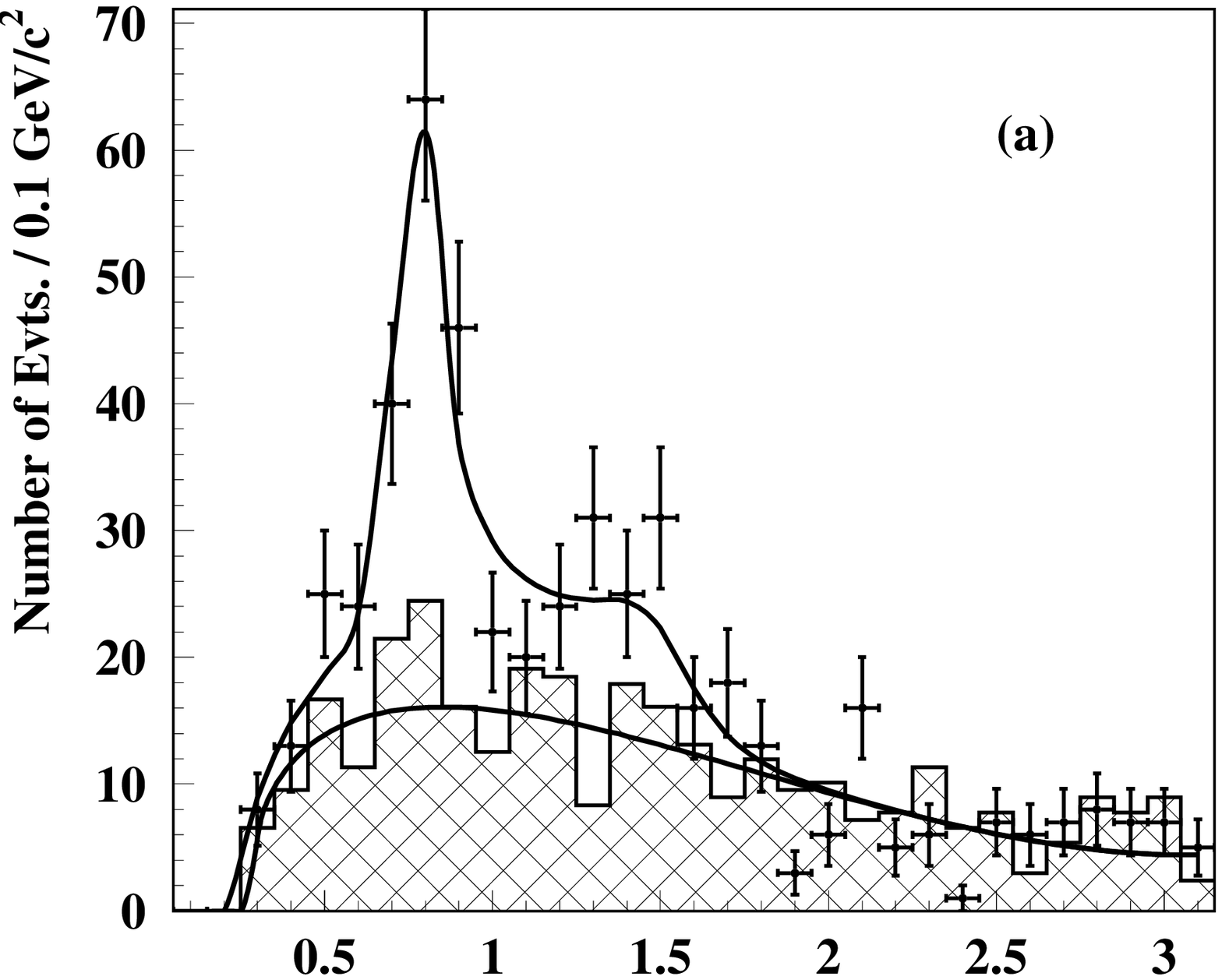}
  \includegraphics[width=0.23\textwidth] {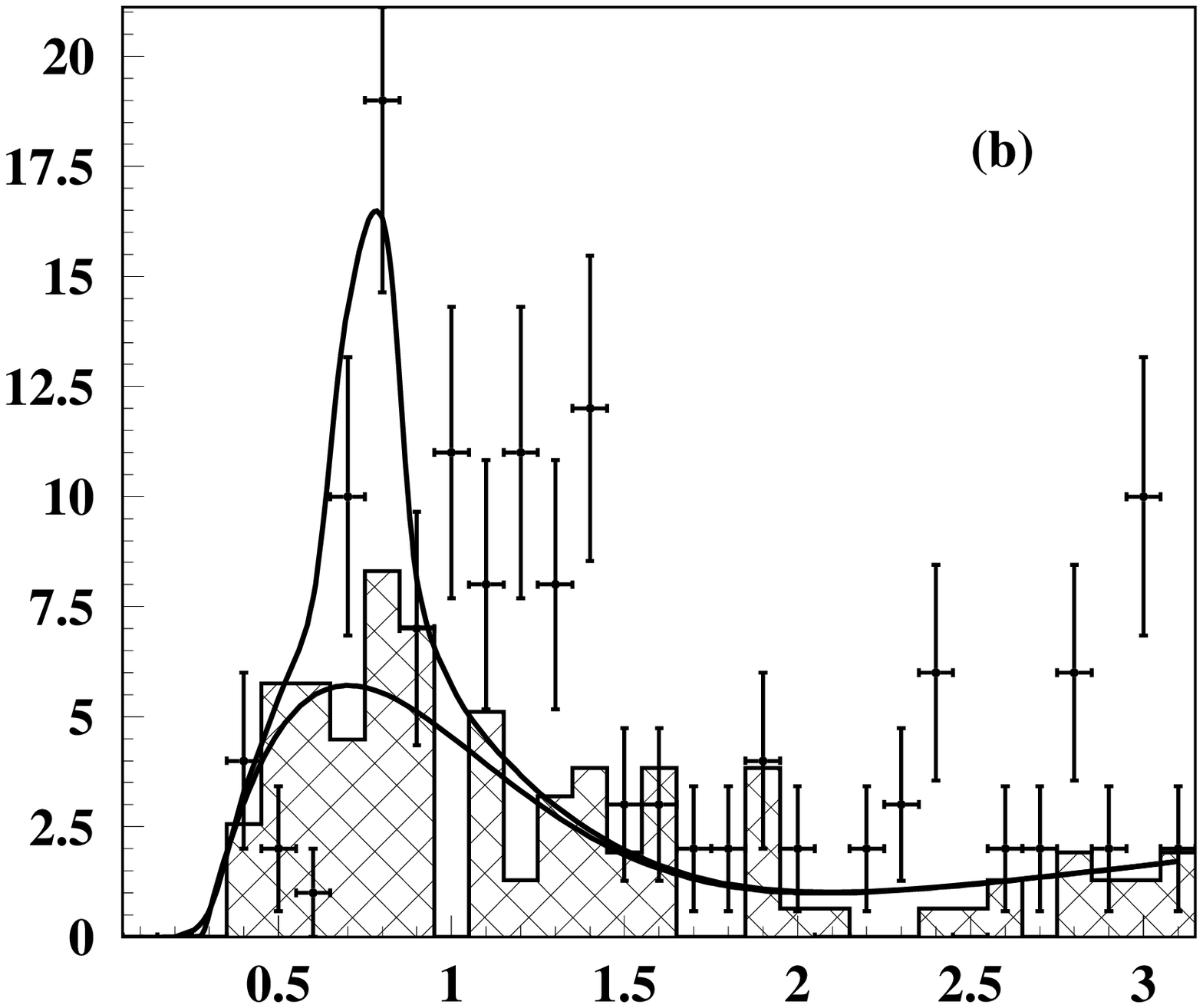}
\vspace*{-1cm}
  \caption{The $\pi^+\pi^-$ invariant mass spectra for the 
    (a) $\bar{B}^0\to D^0\pi^+\pi^-$ and 
    (b) $\bar{B}^{*0}\to D^0\pi^+\pi^-$ candidates.} 
  \label{dpipi_rho}
\end{figure}

Here the Belle results~\cite{d0pipi_belle} on a study of these decays 
are presented. $D^0$ candidates are reconstructed in the 
$\kpi$, $\kpipipi$ and $\kpipin$ final states.
Figure~\ref{dpipi_de} shows the $\de$ distributions
for the $\bar{B}^0\to D^{(*)0}\pi^+\pi^-$ candidates. 
The following branching fractions are measured:
${\cal B}(\bar{B}^0\to D^0\pi^+\pi^-)=(7.5\pm 0.7\pm 1.5)
\times10^{-4}$ and
${\cal B}(\bar{B}^0\to D^{*0}\pi^+\pi^-)=(6.2\pm 1.2\pm 1.7)
\times10^{-4}$.
Figure~\ref{dpipi_rho} shows the $\pi^+\pi^-$ invariant mass
spectra for the selected events. 
These distributions are fitted by a sum of the $\rho^0$,$\sigma$ 
and $f_0(1370)$ resonances. The amplitudes and phases are free
parameters in the fit.
The branching fraction of 
${\cal B}(\bar{B}^0\to D^0\rho^0)=(3.0\pm 1.2\pm 0.4)\times10^{-4}$
and the 90\% CL upper limit 
${\cal B}(\bar{B}^0\to D^{*0}\rho^0)<5.1\times10^{-4}$ are obtained.

\section{$B^-\to D^0 K^{*-}$ (Belle Collaboration)}

The decay $\bdkstar$ can be used for $\phi_3$ determination~\cite{gronau}.
This mode has been previously observed by the CLEO 
Collaboration~\cite{dkstar_cleo}.

$D^0$ mesons are reconstructed in the decay channels $D^0\to\kpi$,
$\kpipipi$ and $\kpipin$ using a $2\sigma$ mass
window from the nominal $D^0$ mass. The $K^{*-}$ candidates are
combined from the $K_S^0\pi^-$ pairs.

\begin{table*}
\caption{The signal yields and branching fractions
for the $\bdkstar$ decay channel.}
\footnotesize
\medskip
\label{dkstar_res}
  \begin{tabular*}{\textwidth}{l@{\extracolsep{\fill}}cccc}\hline\hline
Decay mode  & $\de$ yield & $\mbc$ yield & ${\cal B}$ $(10^{-4})$\\\hline
$D^0\to K^-\pi^+$ & $52.5\pm 8.1$ & $51.6\pm 8.5$ & 
        $6.1\pm 0.9\pm 0.8$\\
$D^0\to K^-\pi^+\pi^0$ & $36.1\pm 6.8$ & $32.9\pm 6.4$ & 
        $5.1\pm 0.9\pm 0.7$\\
$D^0\to K^-\pi^+\pi^+\pi^-$ & $31.0\pm 7.0$ & $29.4\pm 6.8$ & 
        $4.6\pm 1.1\pm 0.8$\\\hline
Weighted mean & & & $5.4\pm 0.6\pm 0.8$\\\hline\hline
  \end{tabular*}
\end{table*}

\begin{figure}
  \includegraphics[width=0.23\textwidth] {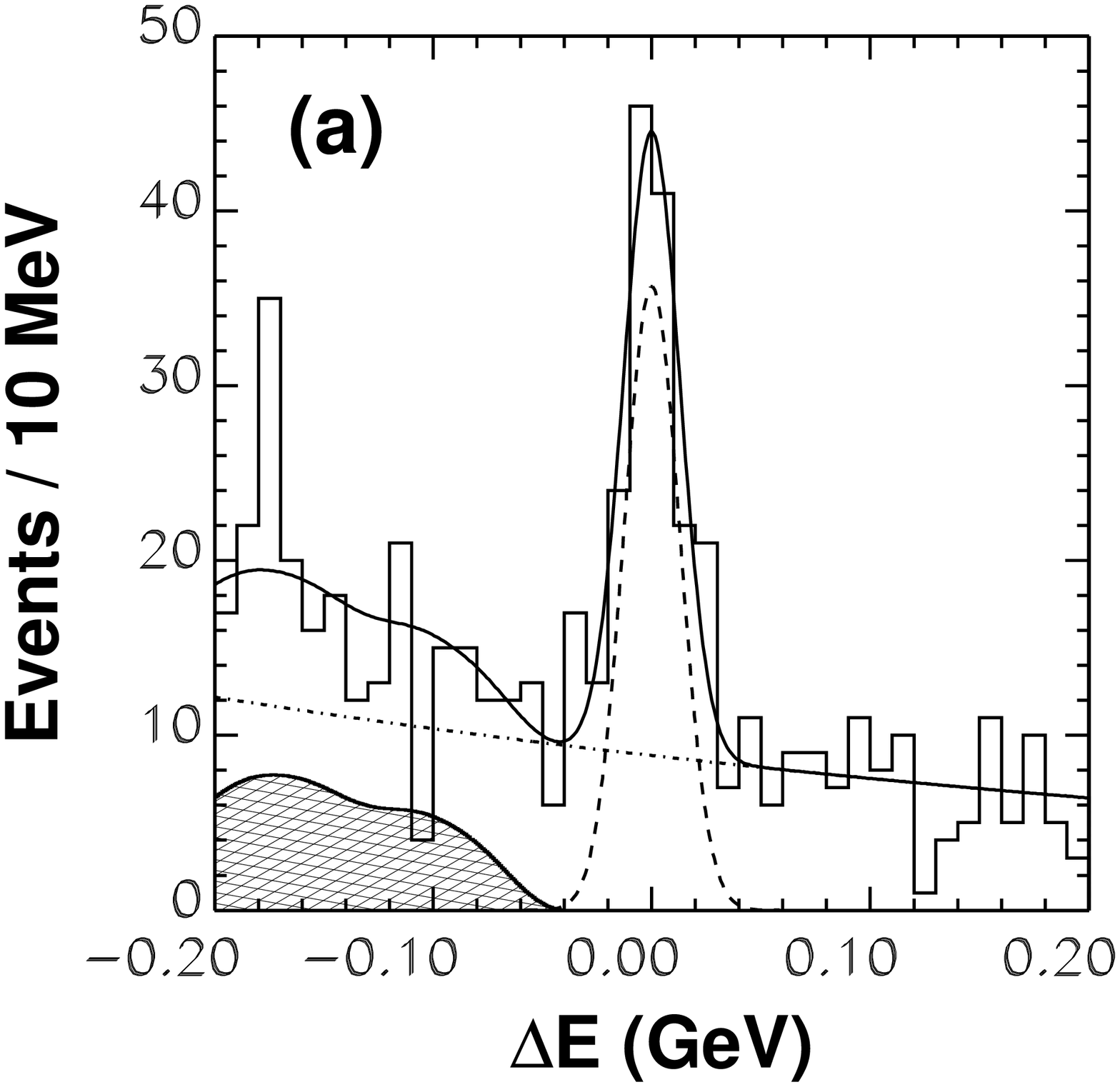}\hfill
  \includegraphics[width=0.23\textwidth] {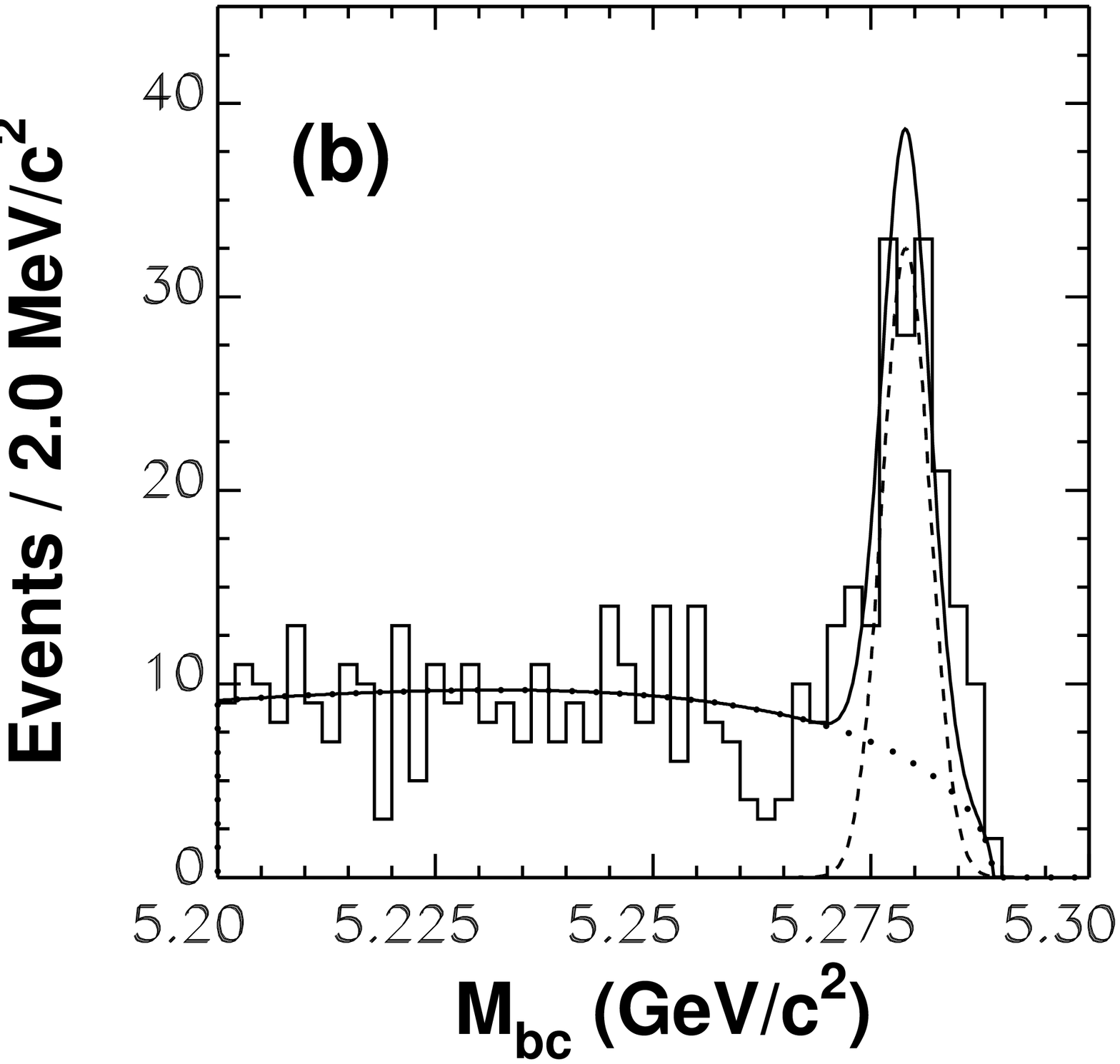}
\vspace*{-1cm}
  \caption{The (a) $\de$, (b) $\mbc$ distributions for the $\bdkstar$ 
    candidates.}
  \label{dkstar_de}
\end{figure}

\begin{figure}
  \includegraphics[width=0.23\textwidth] {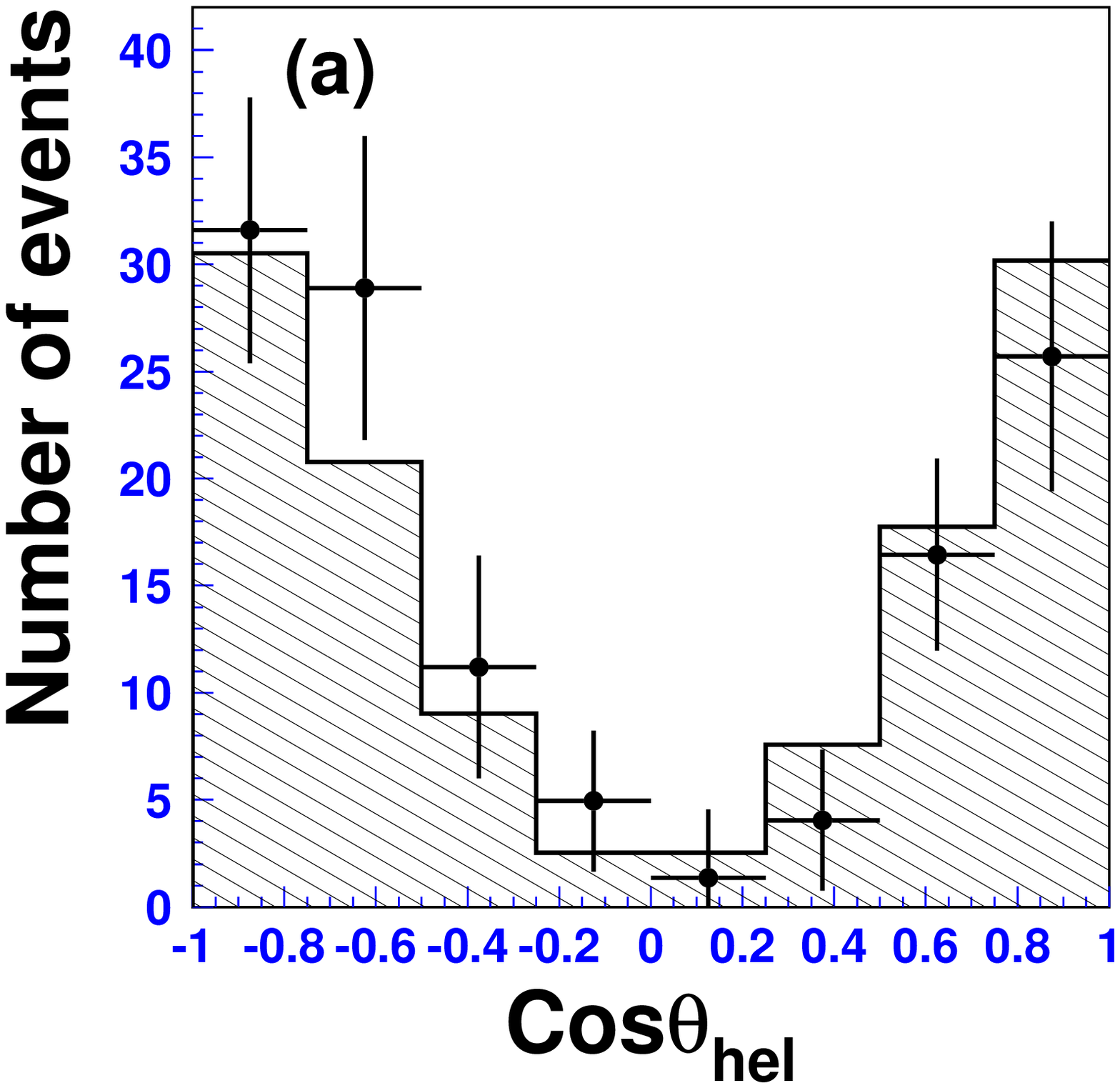}\hfill
  \includegraphics[width=0.23\textwidth] {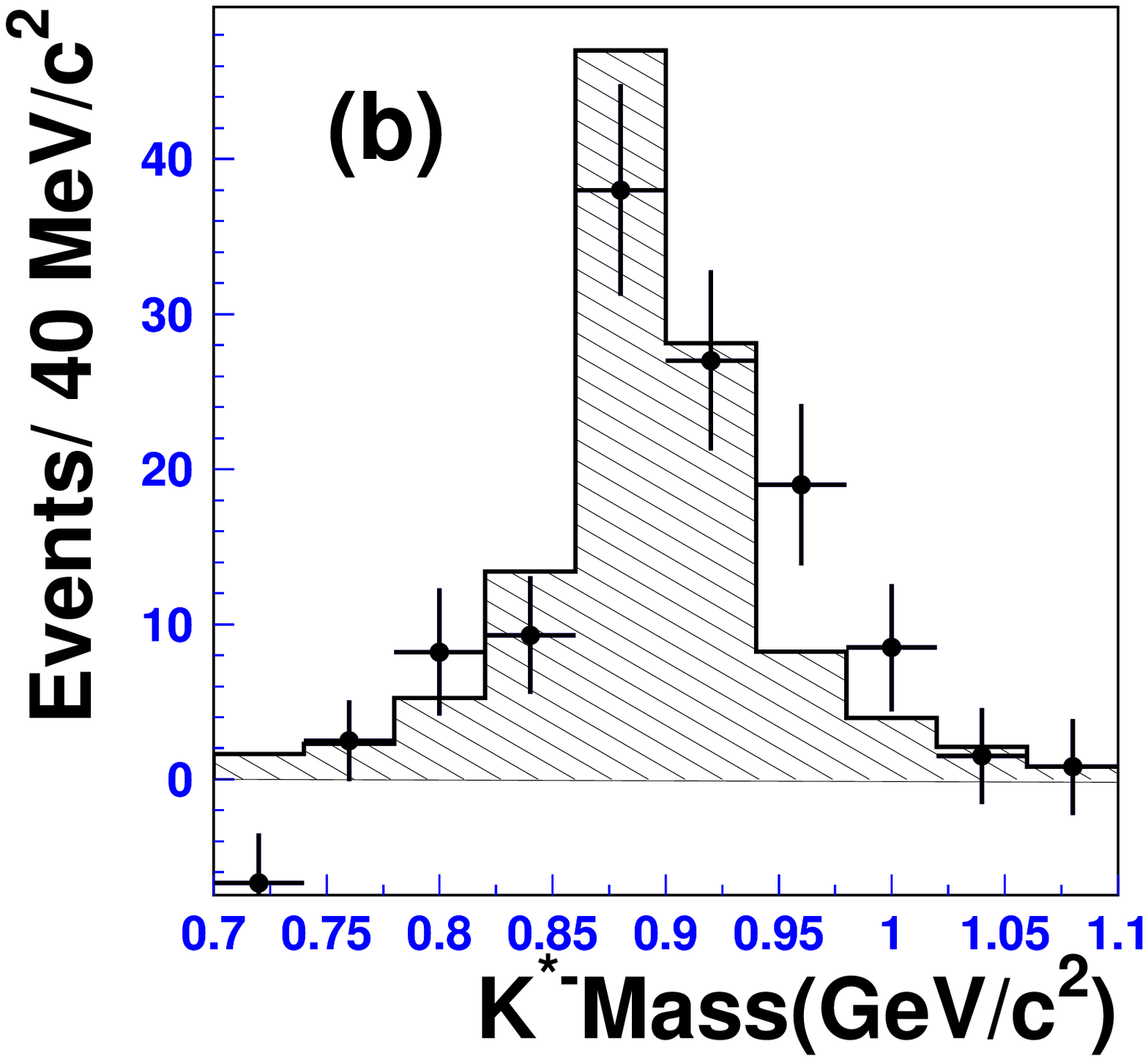}
\vspace*{-1cm}
  \caption{The (a) $K^{*-}$ helicity and (b) $K^{*-}$ invariant mass
    distributions for the $\bdkstar$ signal region.} 
  \label{dkstar_mass}
\end{figure}

The $\de$ and $\mbc$ distributions are presented in
Fig.~\ref{dkstar_de}(a) and (b) respectively. A signal of 
$114.4\pm 13.5$ events with $10.9\sigma$ statistical significance is
observed.
The fit results are presented in Table~\ref{dkstar_res}.
The measured branching fraction
${\cal B}(\bdkstar)=(5.4\pm 0.6\pm 0.8)\times 10^{-4}$
agrees well with the world average~\cite{PDG}.
Figure~\ref{dkstar_mass} shows the $K^{*-}$ helicity and invariant mass 
distributions obtained by fitting the $\mbc$ spectra in each bin.

\section{Conclusion}
A nonzero strong phase $\delta_I$ is obtained using the CLEO and Belle
measurements of the branching fractions ${\cal B}(B\to D\pi)$.
The $2.3\sigma$ difference from zero indicates the presence of final
state interactions in $B\to D\pi$ decays.
The observation of the $\bdsk$ decay by Belle emphasizes the importance
of $W$ exchange or final state rescattering.
In $B^-\to D^{(*)+}\pi^-\pi^-$ decays all four P-wave $D^{**}$ have
been observed and their parameters have been measured by Belle. 
For the broad $D_0^{*0}$ and $D_1^{*0}$ states
this is the first measurement.


\end{document}